\documentclass[a4paper,11pt]{article}
\usepackage{pos}
\usepackage{url}

\title{Galactic Science with the Southern Wide-field Gamma-ray Observatory}
 \ShortTitle{Galactic Science SWGO}

\author*[a,b]{R. L\'opez-Coto}
\author[c]{A. Mitchell}
\author[d]{E. O. Ang{\"u}ner}
\author[e]{G. Giacinti}
\forColl{SWGO}

\affiliation[a]{Istituto Nazionale di Fisica Nucleare, Sezione di Padova, I-35131, Padova, Italy.}
\affiliation[b]{now at Instituto de Astrof\'isica de Andaluc\'ia, CSIC, 18080 Granada, Spain.}
\affiliation[c]{Department of Physics, ETH Zurich, CH-8093 Zurich, Switzerland}
\affiliation[d]{Aix Marseille Univ, CNRS/IN2P3, CPPM, Marseille, France}
\affiliation[e]{Max-Planck-Institut f\"{u}r Kernphysik, Saupfercheckweg 1, 69117 Heidelberg, Germany}

\emailAdd{rlopez@pd.infn.it}

\abstract{The Southern Wide-field Gamma-ray Observatory (SWGO) is a proposed ground-based gamma-ray detector that will be located in the Southern Hemisphere and is currently in its design phase. In this contribution, we will outline the prospects for Galactic science with this Observatory. Particular focus will be given to the detectability of extended sources, such as gamma-ray halos around pulsars; optimisation of the angular resolution to mitigate source confusion between known TeV sources; and studies of the energy resolution and sensitivity required to study the spectral features of PeVatrons at the highest energies. Such a facility will ideally complement contemporaneous observatories in studies of high energy astrophysical processes in our Galaxy.}

\FullConference{37$^{\rm{th}}$ International Cosmic Ray Conference (ICRC 2021)\\
		July 12th -- 23rd, 2021\\
		Online -- Berlin, Germany}

\begin{document}
\maketitle

\section{Introduction}

From the point of inception, a primary focus of SWGO has been its location in the Southern hemisphere, providing access to the Southern sky and the more densely populated regions of the Galactic Plane. As such, Galactic science is a key component of the motivation and scientific agenda for SWGO: a ground-based particle detector in the South, sensitive to very high energy gamma-rays. 
Three key themes push the design and therefore are used for bench-marking SWGO. These are: gamma-ray halos around energetic pulsars; Galactic diffuse gamma-ray emission including the Fermi bubbles; and the search for and study of PeVatrons, accelerators of Galactic Cosmic Rays up to and beyond PeV energies. 

Correspondingly, we explore the constraints that the locations of promising pulsars and halo candidates place on the site location in section \ref{sec:halos}. As the Galactic plane is by nature crowded with sources located at similar positions along the line of sight, particularly along spiral arms, the angular resolution is constrained by the likely level of source confusion from gamma-ray sources in close proximity. Nevertheless, extended gamma-ray sources will lead to inevitable overlap along the line of sight in some cases. 

For studies of low surface brightness Galactic diffuse emission, good background rejection is paramount; SWGO plans to achieve a level that will plausibly enable detection of the Fermi bubbles. In order to detect PeVatrons and study their spectral features the highest energies, such as their spectral curvature, good energy resolution and sensitivity is required (see section \ref{sec:pev}).

Galactic gamma-ray science with SWGO offers a rich opportunity to study the origins of the highest energy Galactic cosmic rays from PeVatrons and the particle transport processes in gamma-ray halos, including particle escape and confinement due to magnetic fields. Additionally, the complex evolutionary history of our Galaxy can be studied through the Fermi bubbles which indicate potential past activity. The ambient sea of Galactic cosmic rays, those which we isotropically detect at Earth, can be probed through studies of the Galactic diffuse gamma-ray emission that arises as a result of interactions with interstellar clouds (producing gamma-rays through the decay of neutral pions) and radiation fields (producing gamma-rays through the leptonic inverse Compton scattering process).

\section{Gamma-ray halos}
\label{sec:halos}
\begin{figure}
\begin{center}
\includegraphics[width=0.75\columnwidth]{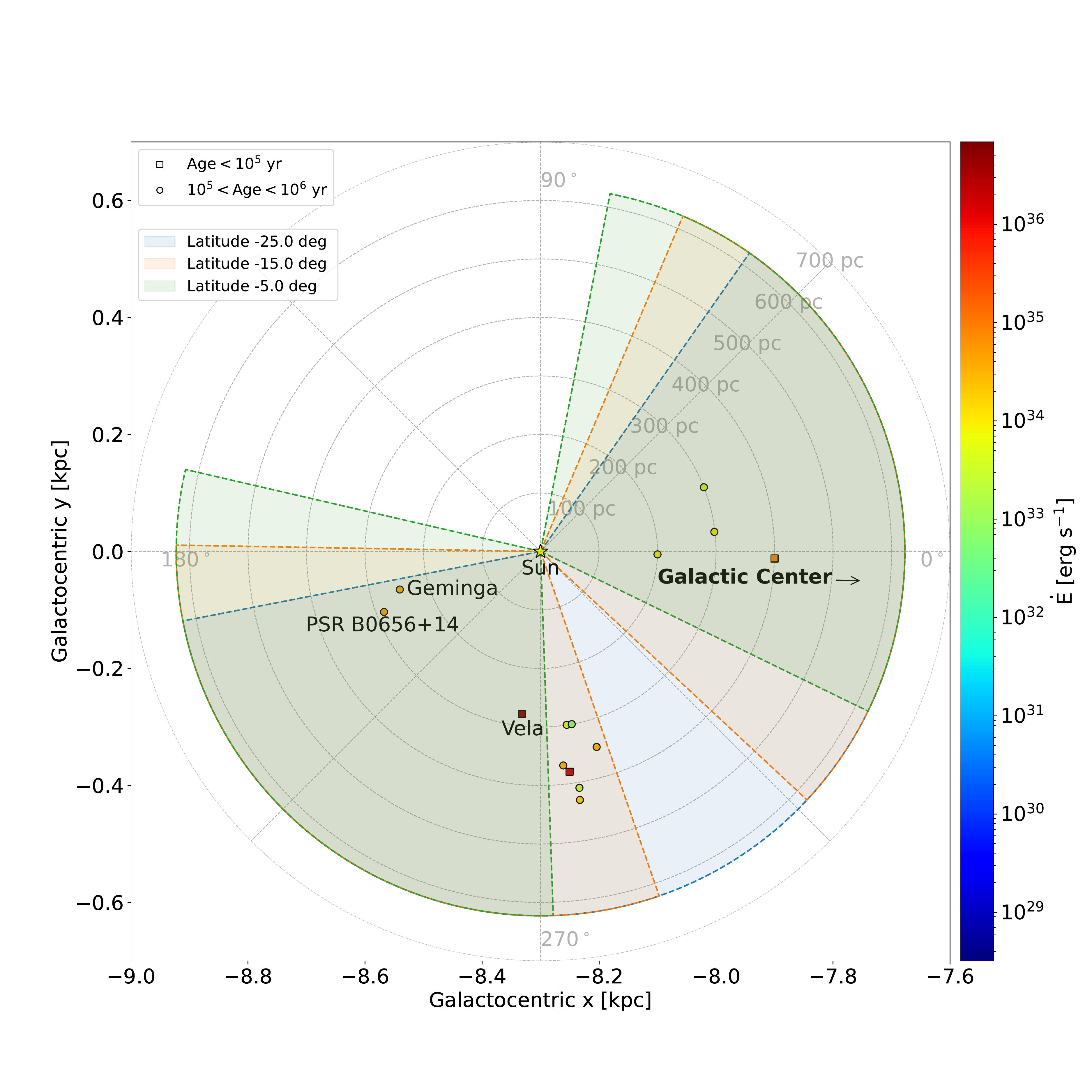}
\caption{Top-down view showing how the observable regions of the Galaxy - and hence the coverage of local ($<500$\,pc), energetic pulsars - depends on the latitude of the chosen observatory site. The outer reach of $\sim 620$\,pc is given by the 5 year sensitivity.}
\label{fig:pulsars_edot}
\end{center}
\end{figure}

New gamma-ray observations in the GeV and TeV domain have revealed a new class of gamma-ray emission regions: the gamma-ray halos \cite{HAWCgeminga}. Gamma-ray halos are characterized by regions in which electrons and positrons generated in the pulsar magnetosphere escape from the Pulsar Wind Nebula and produce a region that is bright in gamma rays. The study of these sources is of paramount importance to unveil acceleration and propagation of Cosmic Rays (CRs) in pulsars and their environments, and calculate the contribution to the local CR fluxes measured by satellites and ground-based detectors. Although Imaging Atmospheric Cherenkov Telescopes (IACTs) are very sensitive to the detection of emission of $\leq 1$ deg, if the central engines generating this emission are very closeby, the source produced at VHE gamma rays extends over regions of several degrees. Wide FoV VHE gamma-ray experiments such as HAWC, LHAASO in the Northern hemisphere have proven to be ideal instruments in the detection and characterization of this kind of extended sources. The future SWGO in the Southern hemisphere is expected to add several detections to this source type and contribute in their understanding. In Fig. \ref{fig:pulsars_edot} we show all the pulsars within 500 pc with an age $<$ 1 Myr and the SWGO reach for its location at different latitudes. We can see that locating it at a latitude of -20$^\circ$, all the pulsars fulfilling the above-mentioned criteria are within reach of the observatory. In this plot we did not take into account the reduction of sensitivity due to a lower exposure of sources culminating at high zenith angle, which will reduce the sensitivity of their observation.

\begin{figure}
\begin{center}
\includegraphics[width=0.95\columnwidth]{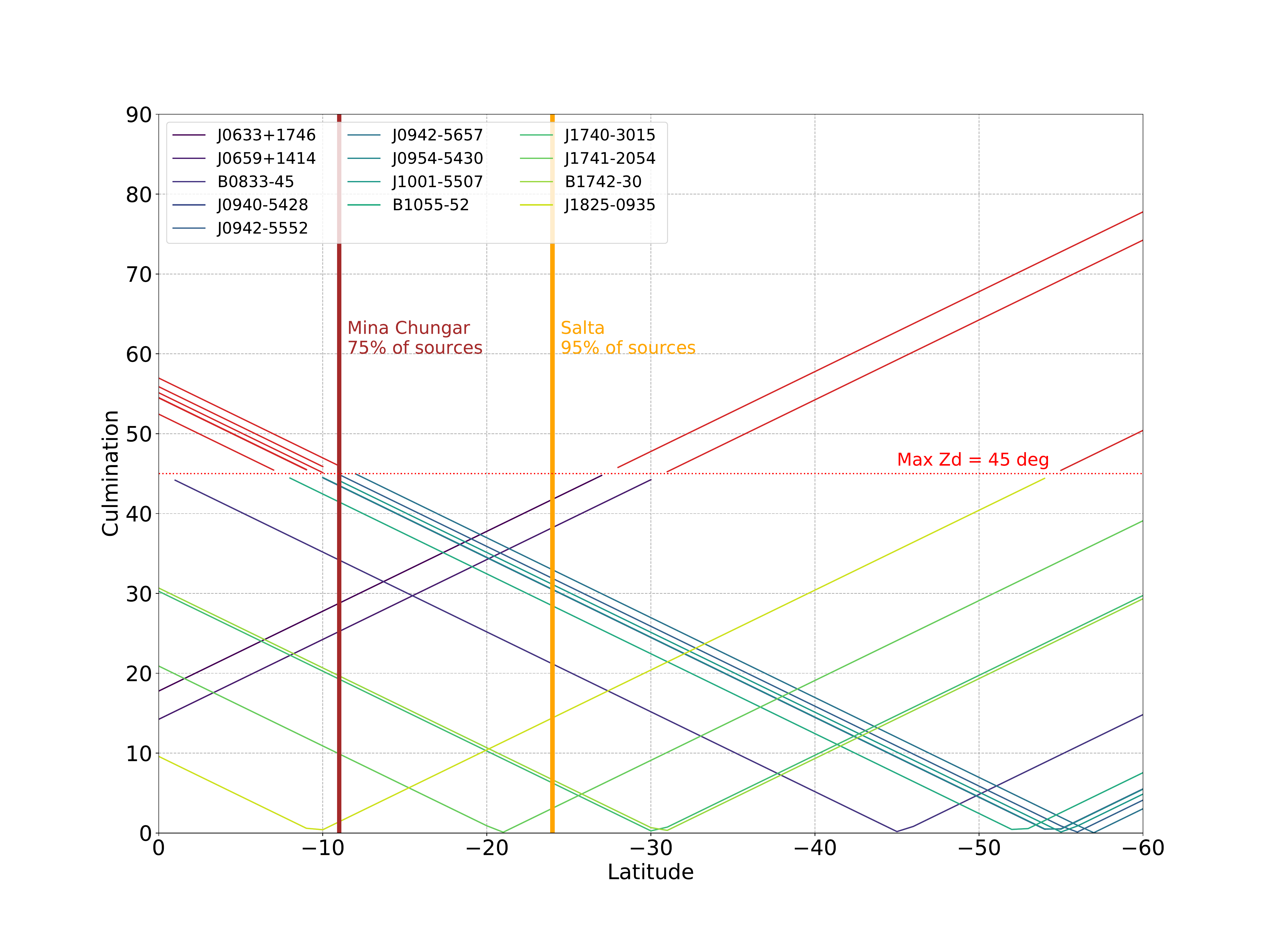}
\caption{The observability of local energetic pulsars as a function of site latitude. Those pulsars for which the diagonal lines, indicating the source culmination on the sky, intersect with the vertical lines for the site latitude are observable from that particular site. Two example site locations are indicated, representing the Northern-most (Mina Chungar) and Southern-most (Salta) candidate sites respectively. } 
\label{fig:tev_halos_observability}
\end{center}
\end{figure}

The source observability seen as their culmination as a function of their declination can be seen in Figure \ref{fig:tev_halos_observability}. We also mark the limits of the Northern-most (Mina Chungar) and Southern-most (Salta) proposed sites. We note that although locating the observatory further South, more sources are accessible, also  known sources like Geminga or PSR B0656+14, important to be able to characterize gamma-ray halo measurements would be barely visible due to their high culmination, and therefore difficult to be detected.

\begin{figure}
\begin{center}
\includegraphics[width=0.9\columnwidth]{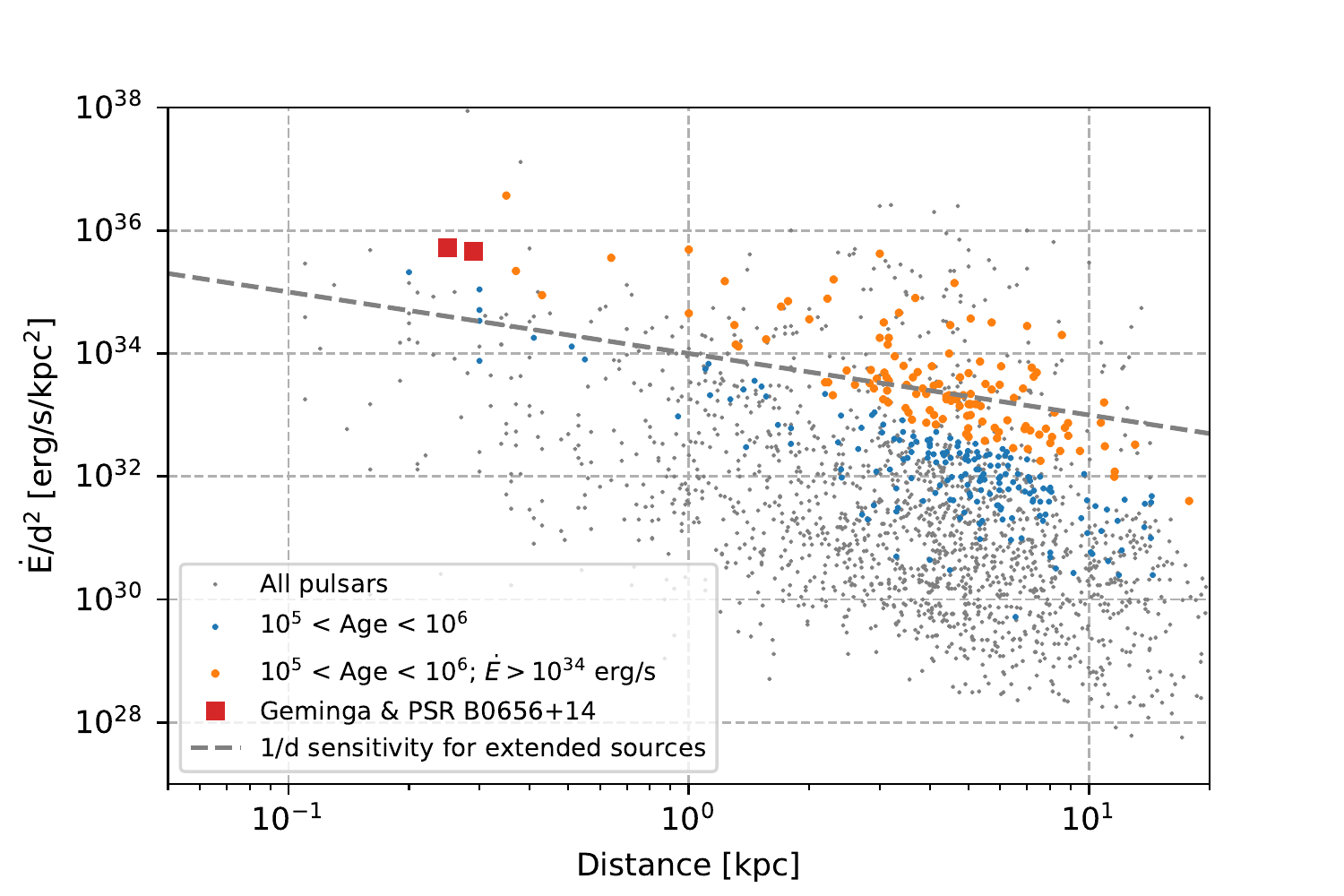}
\caption{Pulsar population from the ATNF catalogue - the most promising halo candidates are shown in orange. The dashed line indicates the SWGO sensitivity to extended sources, assuming similar properties as for the halo around the Geminga pulsar. }
\label{fig:edot_vs_distance}
\end{center}
\end{figure}

Although best to study nearby, very extended sources, SWGO will also be able to reach further away sources with a smaller angular size. In Figure \ref{fig:edot_vs_distance} we include the pulsar population from the ATNF catalogue highlighting the most promising gamma-ray halo candidates. We draw a sensitivity line for SWGO assuming gamma-ray halos to be Geminga-like in terms of size and gamma-ray surface brightness. All sources above the line are candidates to be detected by SWGO. In this plot we did not perform any assumption on the angular resolution of the instrument, that will reduce the sensitivity for the detection of further sources. 

Due to the presence of many sources in the Galactic plane, we will also face problems with source separation. We performed a study using known gamma-ray sources from the H.E.S.S. Galactic Plane Survey (HGPS) and predicted gamma-ray halos \cite{HGPS}. 
As extended emission around pulsars is likely to account for the most numerous source class with the largest radial extents, we modelled the radial evolution hypothetical pulsar wind nebulae and halos for all nearby pulsars listed in the ATNF catalogue \cite{ATNF}. 

The radial evolution of pulsar wind nebulae is modeled in stages, adopting:
\begin{equation}
    R = 1.1\mathrm{pc}\left(\frac{\dot{E_0}}{10^{38}\mathrm{ergs}^{-1}}\right)^{1/5}\left(\frac{t}{10^3\mathrm{yrs}}\right)^{6/5}~,
\end{equation}
from \cite{GaenslerSlane}, for a supernova with a canonical ejection energy of $10^{51}$\,erg and ejecta mass of $10\,\mathrm{M_{sun}}$, then evolving according to:
\begin{equation}
    R(t) \text{  } \propto  \text{  }
    \left\lbrace 
    \begin{array}{ll}
    t^{6/5} & \text{for  } t \leq \tau_0 \\
    t & \text{for  } \tau_0 < t \leq t_{\text{rs}}  \\
    t^{3/10} & \text{for  } t > t_{\text{rs}}. \\
    \end{array}
    \right.
    \label{eq:radiustotal}
\end{equation}
from \cite{PWNpop}. The parameters $\dot{E_0}$ and $\tau_0$ are obtained via $\tau_0=\frac{2\tau_c}{n-1}\left(\frac{P_0}{P}\right)^{n-1}$ and $\dot{E}=\dot{E_0}\left(1+\frac{t}{\tau_0}\right)^{-\frac{n+1}{n-1}}$ where it is assumed that the braking index $n=3$.
Furthermore, we assume that the initial spin period $P_0=30$\,ms and that the reverse shock occurs at $t_{\mathrm{rs}}\sim7$\,kyr. For ages $t>20$\,kyr, we assume that the system then enters the halo evolutionary stage, with the gamma-ray radial extent dominated by the escaped electrons and positrons. During the halo stage, radial evolution is described by $R\propto 2\sqrt{D(E)t}$ where we adopt the diffusion coefficient found by HAWC in the vicinity of Geminga \cite{HAWCgeminga}. Both this model and the HGPS size estimates are valid for the energy range $\sim 1-10$\,TeV. Whilst many assumptions enter into this model, by adopting typical values we intend to capture a likely distribution of radial extents, sufficient to estimate the challenges that will be faced by SWGO in terms of source confusion and required angular resolution.

\begin{figure}
\begin{center}
\includegraphics[width=0.7\columnwidth]{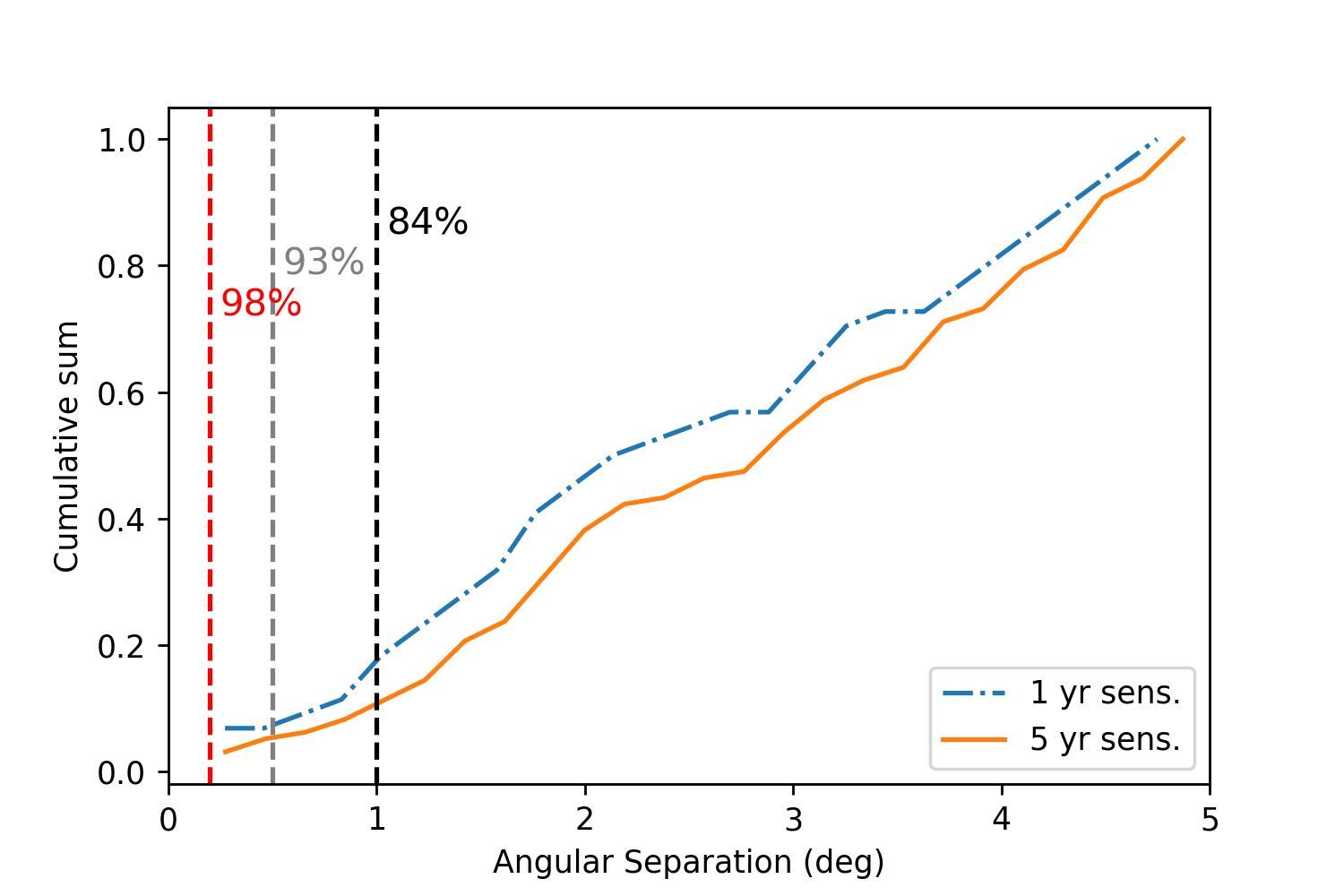}
\caption{The cumulative distribution of angular separations between all known gamma-ray sources in the Southern Galactic plane \cite{HGPS} and predicted pulsar wind nebulae and halos from the ATNF \cite{ATNF}. The angular separation is calculated between the radial extent ($1\sigma$) of the sources. Normalised to the total number of sources with an angular separation $<5^\circ$. Vertical dashed lines indicate the angular resolution required to resolve the indicated percentage of sources. Sources that overlap significantly are not considered.}
\label{fig:separation}
\end{center}
\end{figure}

Combining the known gamma-ray sources and this predicted sample, the angular separation between nearby sources is calculated between the $1\sigma$ radial extents. As it can be drawn from Figure \ref{fig:separation}, we will be able to separate 93\% of the sources if we reach an angular resolution better than 0.5$^\circ$ and 98\% if we reach an angular resolution resolution better than $0.2^\circ$. These values are as a proportion of the total number of sources with angular separations $< 5^\circ$. Sources that are predicted to overlap significantly are not considered, as there will be some inevitable intrinsic overlap along the line of sight that cannot be avoided.

\section{PeVatrons}
\label{sec:pev}

Decades of Cosmic-ray (CR) measurements improved our understanding and helped to construct the energy spectrum of primary CRs observed from Earth. The reconstructed CR spectrum exhibits a remarkable power law in energy over several orders of magnitude. This power law has a break in energy at a few Peta-electronvolts (PeV, $10^{15}$ eV), which is referred to as the knee. Below the knee energies, CRs are believed to be of Galactic origin. Such a spectral feature in the CR spectrum is an indication for the existence of extreme CR factories in our Galaxy, but the sources where they are produced are still unknown. In this context, Galactic sources capable of accelerating particles up to at least PeV energies are called `Galactic PeVatrons'. Understanding the origin of CRs using direct measurements is an impossible task due to the presence of interstellar magnetic fields that deviate the path of charged particles before they reach Earth. Neutral messengers generated by CR interactions, such as gamma rays and neutrinos, are used to determine the origin of CRs. The gamma-ray energy produced in CR interactions is approximately a factor 10 lower than the parent CR energy \cite{Kelner}. Thus, one can probe PeVatrons by searching for gamma-ray emission at energies of 100 TeV and above.

The potential of detecting gamma-rays having energies at/above 100 TeV with a wide field of view, and very high duty cycle, ground-based water Cherenkov telescope arrays has been demonstrated by the current generation instruments HAWC \cite{HAWC100TeV} and LHAASO \cite{LHAASO} located in the Northern hemisphere. Especially, the recent discovery of 12 Galactic sources emitting gamma-rays at energies above 100 TeV, and up to 1.4 PeV \cite{LHAASO}, which are located at Galactic Longitudes $l \geq 10^{\circ}$, provided a robust evidence of the presence of PeVatron sources in our Galaxy. Currently, there is no such instrument, acting as `PeVatron hunter', operational in the Southern hemisphere, which covers the most intriguing part of the Galactic plane in $\gamma$-rays including the Galactic Center region harbouring a PeVatron source \cite{HESSGCPeVatron} and potential E $>$ 100 TeV $\gamma$-ray emission regions such as Westerlund\,1 \cite{Westerlund1} $\&$ HESS\,J1641$-$463 \cite{HESSJ1641} and HESS\,J1702$-$420 \cite{HESSJ1702}. The future SWGO, located in South America, will be able to detect $\gamma$-rays at energies above 100 TeV from this intriguing part of the Galactic Plane. Such future observations will help construct a full picture of the Galactic Plane above 100 TeV $\gamma$-ray energies, therefore allowing population studies of Galactic PeVatrons, which can resolve the 110 year old mystery of the origin of Galactic cosmic-rays. 

\begin{figure}
\begin{center}
\includegraphics[width=1.0\columnwidth]{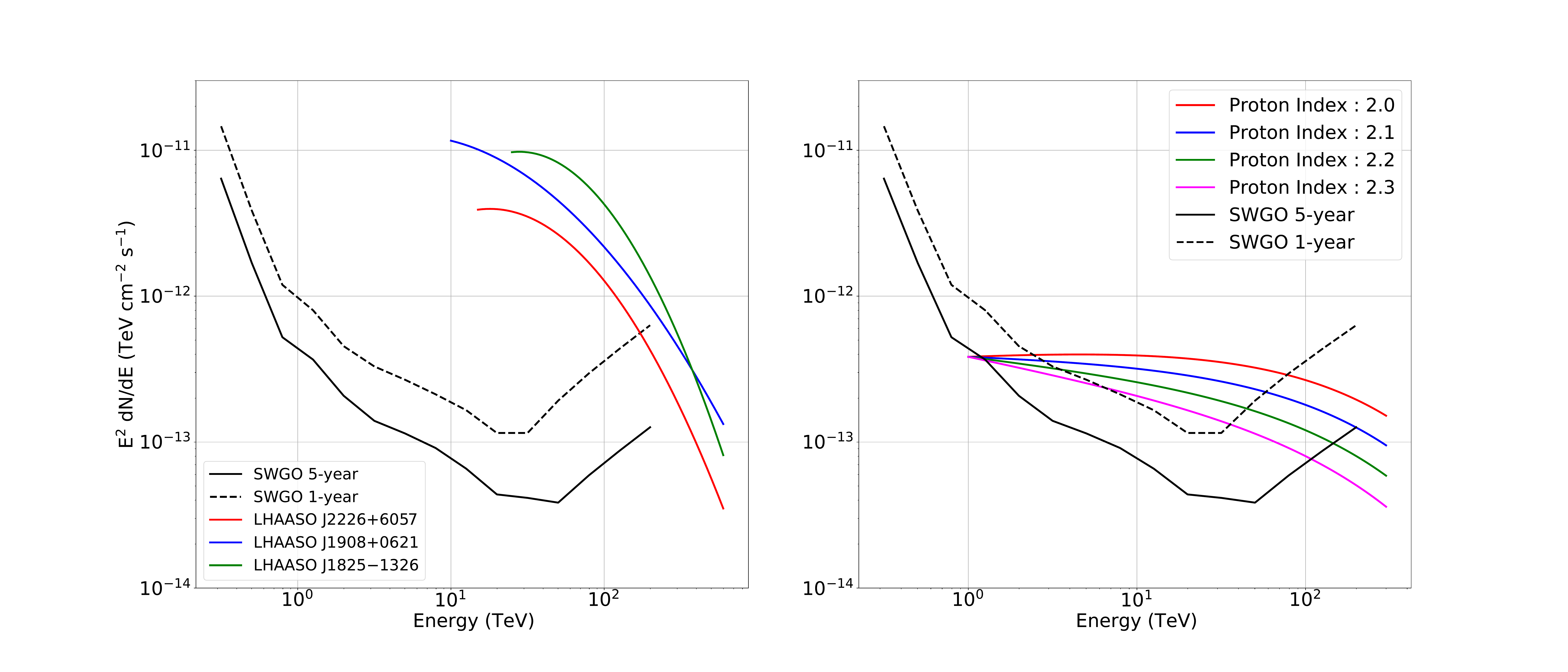}
\caption{Left panel:reconstructed best fit LogParabola models obtained from LHAASO observations of LHAASO\,J2226+6057 (solid-red), LHAASO\,J1908+0621 (solid-blue) and LHAASO\,J1825$-$1326 (solid-green). The LHAASO public data models are taken from http://english.ihep.cas.cn/doc/4035.html. Right panel: $\gamma$-ray emission models from proton-proton interactions followed by subsequent $\pi^{0}$ decay, originated from protons following power-law with exponential cutoff spectral model with fixed proton cutoff energy at 3 PeV, corresponding to the knee feature seen the CR spectrum. Different colors show proton spectra with spectral indices in [2.0, 2.3], while the normalization of proton spectra are arranged to give 5 mCrab $\gamma$-ray flux at 1 TeV, assuming a distance of 4.0 kpc and gas density of 100 cm$^{-3}$. Crab unit is taken as 3.84 $\times$ 10$^{-11}$ cm$^{-2}$ s$^{-1}$ TeV$^{-1}$. The dashed and solid black lines show conservative 1-year and 5-year SWGO sensitivity curves \cite{SWGOwhite}, respectively.}
\label{fig:pevatrons}
\end{center}
\end{figure}

Figure \ref{fig:pevatrons} (left) shows the comparison between conservative 1-year/5-year future SWGO sensitivity curves and reconstructed spectral models from LHAASO observations of three sources showing emission above 100 TeV \cite{LHAASO}, namely LHAASO J2226+6057, LHAASO J1908+0621 and LHAASO J1825$-$1326. It can be seen from the figure that the future SWGO will be able to detect 100 TeV $\gamma$-ray emission from possible similar sources, located at $l \leq 0^{\circ}$, even after $\sim$1-year observations, while deeper 5-year observations will allow detailed spectral studies of such possible sources. Figure \ref{fig:pevatrons} (right) shows the $\gamma$-ray emission, originated from proton-proton interactions, giving rise to 5 mCrab $\gamma$-ray flux (at 1 TeV) level observed from Earth. At least few years deeper observations will be needed in order to reconstruct $\gamma$-ray spectra from such faint Galactic sources. We note that the sensitivity curves given in Fig. \ref{fig:pevatrons} are quite conservative, therefore an improvement in performance is expected. In general, robust detection of PeVatron signature requires detailed spectral investigation not only in $\gamma$-rays but also in the proton parameter space. In this context, derivation of proton spectral cutoff lower limits, together with significant high energy ($\geq$ 100 TeV) $\gamma$-ray spectral flux points would be helpful to understand PeVatron nature of sources. 

\section{Fermi Bubbles and Galactic Diffuse}
\label{sec:bubble}

Bubble-like structures, above and below the Galactic plane have been detected by radio and gamma-ray instruments. The observation of such large structures, together with that of the galactic diffuse emission is particularly challenging for current and even future Imaging Atmospheric Cherenkov Telescopes. To achieve the goal of the detection of Fermi Bubbles and Galactic Diffuse emission, the SWGO observatory plans to optimize the gamma/hadron separation to reach the minimum possible diffuse cosmic-ray residual background level, key for their detection. Preliminary estimates suggest that the reachable levels imply the detectability of the large-scale Galactic Diffuse Emission up to tens of TeV. The detectability of the Fermi Bubbles remains an open question depending on the models extending their emission up to higher energies than those detected by {\it Fermi}-LAT. More detailed calculations are currently ongoing and will be shown in future works.

\section*{Acknowledgements: }
R.L.-C. acknowledges the financial support of the European Union Horizon 2020 research and innovation program under the Marie Sk\l{}odowska-Curie grant agreement No. 754496 - FELLINI. R.L.-C. also acknowledges the financial support from the State Agency for Research of the Spanish MCIU through the ‘Center of Excellence Severo Ochoa’ award to the Instituto de Astrofísica de Andalucía (SEV-2017-0709). The SWGO Collaboration acknowledges the support from the agencies and organizations listed here: \url{https://www.swgo.org/SWGOWiki/doku.php?id=acknowledgements}.

\clearpage
\section*{Full Authors List: SWGO Collaboration}
%
%
\scriptsize
\noindent
P.~Abreu$^1$,
A.~Albert$^2$,
E.\,O.~Angüner$^3$,
C.~Arcaro$^4$,
L.\,H.~Arnaldi$^5$,
J.\,C.~Arteaga-Velázquez$^6$,
P.~Assis$^1$,
A.~Bakalová$^7$,
U.~Barres\,de\,Almeida$^8$,
I.~Batković$^4$
% ...
\vspace{1cm}

\noindent
$^{1}$ {Laboratório de Instrumentação e Física Experimental de Partículas (LIP), Av. Prof. Gama Pinto 2, 1649-003 Lisboa, Portugal\\}
$^{2}$ {Physics Division, Los Alamos National Laboratory, P.O. Box 1663, Los Alamos, NM 87545, United States\\}
$^{3}$ {Aix Marseille Univ, CNRS/IN2P3, CPPM, 163 avenue de Luminy - Case 902, 13288 Marseille cedex 09, France\\}
$^{4}$ {University of Padova, Department of Physics and Astronomy \& INFN Padova, Via Marzolo 8 - 35131 Padova, Italy\\}
$^{5}$ {Centro Atómico Bariloche, Comisión Nacional de Energía Atómica, S. C. de Bariloche (8400), RN, Argentina\\}
$^{6}$ {Universidad Michoacana de San Nicolás de Hidalgo, Calle de Santiago Tapia 403, Centro, 58000 Morelia, Mich., México\\}
$^{7}$ {FZU, Institute of Physics of the Czech Academy of Sciences, Na Slovance 1999/2, 182 00 Praha 8, Czech Republic \\}
$^{8}$ {Centro Brasileiro de Pesquisas Físicas, R. Dr. Xavier Sigaud, 150 - Rio de Janeiro - RJ, 22290-180, Brazil\\}
$^{9}$ {Academic Department of Physics – Faculty of Sciences – Universidad Nacional de San Antonio Abad del Cusco (UNSAAC), Av. de la Cultura, 733, Pabellón C-358, Cusco, Peru\\}
$^{10}$ {Instituto de Física, Universidad Nacional Autónoma de México, Sendero Bicipuma, C.U., Coyoacán, 04510 Ciudad de México, CDMX, México \\}
$^{11}$ {Dipartimento di Fisica, Università degli Studi di Torino, Via Pietro Giuria 1, 10125, Torino, Italy\\}
$^{12}$ {RWTH Aachen University, Physics Institute 3, Otto-Blumenthal-Straße, 52074 Aachen, Germany \\}
$^{13}$ {University of Maryland, College Park, MD 20742, United States\\}
$^{14}$ {Durham University, Stockton Road, Durham, DH1 3LE, United Kingdom\\}
$^{15}$ {Astronomical Observatory, University of Warsaw, Aleje Ujazdowskie 4, 00478 Warsaw, Poland\\}
$^{16}$ {Facultad de Ciencias en Física y Matemáticas UNACH, Boulevard Belisario Domínguez, Km. 1081, Sin Número, Terán, Tuxtla Gutiérrez, Chiapas, México\\}
$^{17}$ {Facultad de Ciencias Exactas, Universidad Nacional de Salta, Avda. Bolivia N° 5150, (4400) Salta Capital, Argentina\\}
$^{18}$ {Department of Aerospace Science and Technology, Politecnico di Milano, Via Privata Giuseppe La Masa, 34, 20156 Milano MI, Italy\\}
$^{19}$ {Facultad de Ciencias Físico Matemáticas, Benemérita Universidad Autónoma de Puebla, C.P. 72592, México\\}
$^{20}$ {Instituto de Astronomia y Fisica del Espacio (IAFE, CONICET-UBA), Casilla de Correo 67 - Suc. 28 (C1428ZAA), Ciudad Autónoma de Buenos Aires, Argentina\\}
$^{21}$ {Universidad de Guadalajara, Blvd. Gral. Marcelino García Barragán 1421, Olímpica, 44430 Guadalajara, Jal., México\\}
$^{22}$ {Instituto de Física de São Carlos, Universidade de São Paulo, Avenida Trabalhador São-carlense, nº 400, Parque Arnold Schimidt - CEP 13566-590, São Carlos - São Paulo - Brasil\\}
$^{23}$ {INFN - Roma Tor Vergata and INAF-IAPS, Via del Fosso del Cavaliere, 100, 00133 Roma RM, Italy\\}
$^{24}$ {Dept. of Physics and CCTVal, Universidad Tecnica Federico Santa Maria, Avenida España 1680, Valparaíso, Chile\\}
$^{25}$ {Universität Würzburg, Institut für Theoretische Physik und Astrophysik, Emil-Fischer-Str. 31, 97074 Würzburg, Germany\\}
$^{26}$ {Department of Physics, and the Wisconsin IceCube Particle Astrophysics Center (WIPAC), University of Wisconsin, 222 West Washington Ave., Suite 500, Madison, WI 53703, United States\\}
$^{27}$ {TUBITAK Space Technologies Research Institute, ODTU Campus, 06800, Ankara, Turkey\\}
$^{28}$ {Friedrich-Alexander-Universität Erlangen-Nürnberg, Erlangen Centre for Astroparticle Physics, Erwin-Rommel-Str. 1, D 91058 Erlangen, Germany\\}
$^{29}$ {Tecnologico de Monterrey, Escuela de Ingeniería y Ciencias, Ave. Eugenio Garza Sada 2501, Monterrey, N.L., 64849, México\\}
$^{30}$ {Max-Planck-Institut f\"ur Kernphysik, P.O. Box 103980, D 69029 Heidelberg, Germany\\}
$^{31}$ {Università di Napoli “Federico II”, Dipartimento di Fisica “Ettore Pancini”,  and INFN Napoli, Complesso Universitario di Monte Sant'Angelo - Via Cinthia, 21 - 80126 - Napoli, Italy \\}
$^{32}$ {University of Granada, Campus Universitario de Cartuja, Calle Prof. Vicente Callao, 3, 18011 Granada, Spain\\}
$^{33}$ {IFLP, Universidad Nacional de La Plata and CONICET, Diagonal 113, Casco Urbano, B1900 La Plata, Provincia de Buenos Aires, Argentina\\}
$^{34}$ {University of Utah, 201 Presidents' Cir, Salt Lake City, UT 84112, United States\\}
$^{35}$ {Michigan Technological University, 1400 Townsend Drive, Houghton, MI 49931, United States\\}
$^{36}$ {Dipartimento di Fisica e Astronomia "E. Majorana", Catania University and INFN, Catania, Italy\\}
$^{37}$ {APC--IN2P3/CNRS, Université de Paris, Bâtiment Condorcet, 10 rue A.Domon et Léonie Duquet, 75205 PARIS CEDEX 13, France\\}
$^{38}$ {University of Leicester, University Road, Leicester LE1 7RH, United Kingdom\\}
$^{39}$ {Department of Physics, University of Trieste and INFN Trieste, via Valerio 2, I-34127, Trieste, Italy \\}
$^{40}${Centro de Investigación en Computación, Instituto Politécnico Nacional, Av. Juan de Dios Bátiz S/N, Nueva Industrial Vallejo, Gustavo A. Madero, 07738 Ciudad de México, CDMX, México\\}
$^{41}$ {Department of Physics and Mathematics, Universidad de Monterrey, Av. Morones Prieto 4500, San Pedro Garza García 66238, N.L., México\\}
$^{42}$ {Department of Physics and Astronomy, University of Alabama, Gallalee Hall, Tuscaloosa, AL 35401, United States\\}
$^{43}$ {Instituto de Tecnologías en Detección y Astropartículas (CNEA-CONICET-UNSAM), Av. Gral Paz 1499 - San Martín - Pcia. de Buenos Aires, Argentina\\}
$^{44}$ {Department of Physics, ETH Zurich, CH-8093 Zurich, Switzerland\\}
$^{45}$ {Instituto de Ciencias Nucleares, Universidad Nacional Autónoma de México (ICN-UNAM), Cto. Exterior S/N, C.U., Coyoacán, 04510 Ciudad de México, CDMX, México\\}
$^{46}$ {Departamento de Física, Facultad de Ciencias Básicas, Universidad Metropolitana de Ciencias de la Educación, Av. José Pedro Alessandri 774, Ñuñoa, Santiago, Chile\\}
$^{47}$ {Department of Physics, University of Seoul, 163 Seoulsiripdaero, Dongdaemun-gu, Seoul 02504, Republic of Korea \\}
$^{48}$ {Institut de recherche sur les lois fondamentales de l'Univers (IRFU), CEA, Université Paris-Saclay, F-91191 Gif-sur-Yvette, France\\}
$^{49}$ {Laboratoire Univers et Particules de Montpellier, CNRS,  Université de Montpelleir, F-34090 Montpellier, France\\}
$^{50}$ {Instituto Nacional de Astrofísica, Óptica y Electrónica (INAOE), Luis Enrique Erro 1, Puebla, México\\}

%
%
%

\end{document}